\documentclass[manuscript]{aastex}

\usepackage{graphicx}
\usepackage{subfigure}
\usepackage{xcolor}
\usepackage{mathtext,bbm,amsmath,amsfonts,amssymb,indentfirst,syntonly,graphicx}
\usepackage{mathtools}
\usepackage{slashbox}
\usepackage[english]{babel}
\usepackage{calc}
\usepackage{tikz}
\usepackage{epsfig}
\usepackage{multirow}

\def \nustar {\emph{NuSTAR}}
\def \rxte {\emph{RXTE}}

\shorttitle{The NuSTAR View of a QPO Evolution of GRS 1915+105}
\shortauthors{Zhang et al.}

\begin{document}


\title{The \emph{NuSTAR} View of a QPO Evolution of GRS 1915+105}


\author{Liang Zhang\altaffilmark{1} and Li Chen\altaffilmark{1}}
\affil{Department of Astronomy, Beijing Normal University, Beijing 100875, China}
\email{201431160006@mail.bnu.edu.cn, chenli@bnu.edu.cn}

\author{Jin-lu Qu\altaffilmark{2}}
\affil{Laboratory for Particle Astrophysics, CAS, Beijing 100049, China}
\email{qujl@ihep.ac.cn}

\and

\author{Qing-cui Bu\altaffilmark{1}}
\affil{Department of Astronomy, Beijing Normal University, Beijing 100875, China}
\email{buqc@mail.bnu.edu.cn}




\begin{abstract}
  We report a timing analysis of the black hole binary GRS 1915+105 with the \nustar\ observatory. A strong type-C QPO below 2 Hz appears in the power density spectrum during the whole observation, whose frequency is correlated with the 3-25 keV count rate. The QPO shows a sudden increase in frequency along with an increase in flux and a softening of the spectrum. We discuss the possible origin of the QPO and the reasons that lead to the QPO frequency variation. It is suggested that the reflection component has little influence on QPO frequency and the increase in QPO frequency could be associated with the inward motion of the outer part of the disk.
\end{abstract}





\section{Introduction}

Most black hole low-mass X-ray binaries are transient sources (black hole transients, BHTs) that are characterized by short outbursts after long periods in the quiescent state. During the outburst, four main spectral states have been identified according to the spectral and timing properties, viz. low-hard state (LHS), hard-intermediate state (HIMS), soft-intermediate state (SIMS) and high-soft state (HSS), corresponding to different branches in the hardness-intensity diagram (HID) (see \citet{Belloni10} and \citet{Belloni11} for a recent reviews). The energy spectra of BHTs usually consist of two components: a thermal component associated with an optically thick accretion disk and a power-law component attributed to Comptonization of soft photons in a hot corona. In some high inclination systems, an additional disk reflection component is required in the spectral model \citep{Done01}. The hard-soft spectral evolution can be explained by truncated disk model \citep{Done07}, assuming that the outer geometrically thin, optically thick disk is truncated at some radius larger than the last stable orbit. The inner accretion flow is a hot, geometrically thick, optically thin flow. The inward movement of the truncation radius with increasing mass accretion rate leads to a softer spectrum. Low frequency quasi-period oscillations (LF-QPOs) with centroid frequencies ranging from a few mHz to 20 Hz have been observed in most BHTs and are thought to originate in the innermost accretion flows \citep{Motta11}. Therefore, studying the QPO properties helps us to understand the accretion physics around black holes. The LF-QPOs in BHTs can be classified into three types, named type-C, type-B and type-A, based on the quality factor $Q$, fractional root-mean-square (rms), noise component and phase lag \citep{Casella04,Casella05}. Different types of QPOs relate to different spectral states.

GRS 1915+105 is a well-studied BHT and has been continuous outburst for 20 years since it is discovered for the first time in 1992 \citep{Remillard06}. This famous system shows complex timing and spectral properties that are different from other BHTs. \citet{Belloni00} classified the X-ray variability patterns into 12 separate classes based on their light curves and CCDs characteristics, while each of the different types of variabilities can be reduced to three basic states, dubbed state A, state B and state C (see Figure 1). State C is seen in a hard state with a disappearing inner accretion disk. State B and state A are two softer states corresponding to different temperatures of inner disk. The $\chi$ class, which is only found in state C, shows strong type-C QPOs with variable frequency ranging from 0.5 to 10 Hz. It is suggested that these LF-QPOs are linked to the properties of accretion disk since their centroid frequencies are correlated with the disk flux \citep{Markwardt99}. However, the energy dependence of the QPO rms amplitudes implied that the hard component may play an important role in their origins \citep{Belloni11}. These results give us a hint that such QPOs may be generated in the coupled regions between thermal and hard components. So far, one of the most promising models for type-C QPOs is Lense-Thirring precession of the inner hot flow \citep{Ingram09,Ingram11}, and the variable frequency could be interpreted as the movement of the truncation radius.

The timing properties, especially phase lag behaviors of the 0.5-10 Hz QPOs observed in GRS 1915+105 have been widely studied. \citet{Reig00} found that there exists a strong anti-correlation between phase lag and QPO frequency, and the phase lag switches from positive to negative as the QPO frequency increases above 2 Hz. The relation was also studied by \citet{Pahari13}. According to the phase lags at the QPO fundamental and first harmonic, the 0.5-10 Hz QPOs can be further divided into three groups: (1) 0.5-2.0 Hz QPOs, have positive lags at both the QPO fundamental and first harmonic frequencies; (2) 2.0-4.5 Hz QPOs, have negative lags at the fundamental frequencies, while positive lags at the first harmonic frequencies; (3) 4.5-10 Hz QPOs, have no harmonic peaks \citep{Lin00}. These studies suggest that the QPOs above 2 Hz and below 2 Hz may arise from different regions or relate to different physical processes.

The \emph{Nuclear Spectroscopic Telescope Array} (\emph{NuSTAR}), launched on 2012 June 13, consists of two co-aligned hard X-ray grazing incidence telescopes \citep{Harrison13}. Unlike an integrating CCD, \nustar\ has a triggered readout without photon pile-up. On this point, it is suitable for timing analysis, and could help us better understand the origins of different types of QPOs. Due to the broad energy band (3-79 keV) and unprecedented sensitivity in the hard X-ray band, \nustar\ is fit to study the high-energy radiation mechanism and the QPO properties in a higher energy band. \nustar\ observed GRS 1915+105 on 2012 July 3. A detailed spectral analysis was reported by \citet{Miller13} (here after M13), but the timing properties have not been studied. In this paper, we focus on studying the timing properties of this observation. We find a QPO exhibit a sudden increase in centroid frequency along with a flux increase. During the transition, the spectral properties have changed.

\section{OBSERVATION AND DATA REDUCTION}

 GRS 1915+105 was observed with \nustar\ on 2012 July 3 from 1:24:21 to 18:01:17 UT with net exposure times of 14.7 ks and 15.2 ks for the focal plane modules A and B (FPMA and FPMB), respectively.  We reduced and analyzed the data using NuSTARDAS version 1.3.1 and CALDB version 20131223 in conjunction with FTOOLS 6.15. A standard data filtering criteria was used with the NUPIPELINE task\footnote{Please see the \nustar\ Data Analysis Software Guide for a detailed description of the data processing.}. Spectra and light curves were extracted from a $90''$ circle centered on GRS 1915+105. Background spectra were extracted from a circle of equivalent radius close to the source region. The background-subtracted spectra were fitted using XSPEC version 12.7 in the energy range 3-79 keV and a systematic error of $0.6~\%$ was added to the spectra.

\section{DATA ANALYSIS AND RESULTS}

\subsection{Timing Analysis}

For our timing analysis, we extracted 3-25 keV light curves from both FPMA and FPMB. The light curves were not background-subtracted because the influence of the background is negligible. For each of the two detectors, we divided the data into 16 s segments and calculated a power density spectrum (PDS) for each of them. We use a time resolution of 1/32 s because we are interested in the 0.5-10 Hz QPOs (note the Nyquist frequency is 16 Hz). The resulting PDSs were then averaged using a logarithmic rebinning. The Leahy normalization \citep{Leahy83} was used without subtracting white noise. Note the PDS has artifacts at multiples of 1 Hz\footnote{Please see the Q16 on the \nustar\ Frequently Asked Questions for a detailed explanation of this problem (http://heasarc.gsfc.nasa.gov/docs/nustar/nustar\_faq.html).}. In our analysis, we removed these artifacts since they have no influence on our results. We fitted the PDSs using a model consisting of a constant and multiple Lorentzians.

The FPMA light curve is shown in Figure 2 (top panel) for the 3-25 keV energy band. The observation exhibits moderate variability without any flares and dips. It belongs to the ``plateau" state or class $\chi$ of state C in the classification by \citet{Belloni00}. From Figure 2, we can see a sudden increase in count rate from $\sim250$ cts/s to $\sim270$ cts/s that occurred near 40 ks after the start of the observation. In order to check which energy band dominates the transition in count rate, we plot the 3-9 keV, 9-25 keV light curves together with hardness (9-25 keV/3-9 keV) in Figure 3. It is evident that the increase in count rate is observed in the 3-9 keV energy band, however not in the 9-25 keV light curve. The count rate in the 9-25 keV band nearly remains constant, which leads to a drop in hardness near 40 ks.

The timing properties of the $\chi$ class are characterized by 0.5-10 Hz QPOs. We observe a strong QPO at $1.54\pm0.01$ Hz that is present in the PDS of the entire observation. To check the variation of the QPO, we plot the dynamical power spectrum in Figure 2 (bottom). We find that the frequency of the QPO shows complex variability: it almost keeps a constant value of $\sim1.5$ Hz before 40 ks, however, sharply increases to $\sim1.9$ Hz at the time of the transition in count rate. For detailed investigations, we split the data into two intervals, dubbed interval 1 and interval 2, using the time of 40 ks as the divider. The power density spectra for the two intervals are shown in Figure 4. Note that we only show the results of FPMA in this paper since the timing properties of FPMB are in accord with FPMA. In both cases, a strong QPO and its second harmonic are observed, together with a subharmonic peak and a band-limited noise (BLN) component. The values of centroid frequency, fractional rms amplitude and quality factor of the fundamental QPO and its harmonics are listed in Table 1. Comparing interval 2 to interval 1, the centroid frequency and rms of the fundamental QPO increase by $\sim22~\%$ and $\sim8~\%$, respectively, but the coherence (quality factor) is lower. The centroid frequencies of the second harmonic and subharmonic components also have a small increase.

From inspection of the dynamical power spectrum and the light curve, we find that the QPO frequency depends on the count rate to some extent. Both of the QPO frequency and 3-25 keV count rate have no significant change in interval 1, while they slowly decrease to a minimum value at the beginning of interval 2 and then increase. In order to study further the relation between the QPO frequency and the source flux, we divided the observation into several segments and calculated a PDS for each of them. The result is shown in Figure 5. The QPO centroid frequency is linearly correlated with the source count rate. The correlation is consistent with \rxte\ observation reported by \citet{Markwardt99}.

We also studied the energy dependence of the QPO rms amplitude. For QPOs in both interval 1 and 2, the rms amplitude increases with photon energy below 10 keV and then flattens above 10 keV (see Figure 6).

\subsection{Spectral Analysis}

A spectral analysis of the whole \nustar\ observation was made by \citet{Miller13}. They found that the spectrum is best-fitted with a relativistically blurred disk reflection model. This model has a form of constant~$\times$~tbabs~((kerrconv~$\times$~reflionx)~+~cutoffpl) in XSPEC. The FPMA and FPMB spectra were fitted simultaneously using a constant to reflect absolute flux levels of the two detectors. The interstellar absorption was fitted with the ``tbabs". ``Reflionx" describes the reflection spectrum emitted by an ionized accretion disk of constant density \citep{Ross05}. The reflection spectrum is convolved with the effects of relativistic smearing via ``kerrconv" \citep{Brenneman06} (see M13 for a detailed description of the model). Following their method, we redid spectral fits and our results are consistent with theirs. In order to check spectral difference between interval 1 and 2, we carried out a spectral analysis for each of them over the 3-79 keV band using the best-fit model from M13. The interstellar absorption was fixed to $6.5\times10^{22}~\rm{cm}^{-2}$, which is close to the \rxte\/PCA analysis by \citet{Belloni97} and the inner disk inclination was fixed to $70^{\circ}$ \citep{Mirabel94,Rodriguez99}. GRS 1915+105 is known to contain a rapidly rotating (Kerr) black hole. \citet{McClintock06} found that the dimensionless spin parameter of GRS 1915+105 is above 0.98 depending on how the thermal continuum is fitted. In our fits, the spin was fixed to 0.98. The abundance of Fe was fixed to the solar value. Notice that the above parameter values are all close to the best-fit results in M13. The reason why we fixed these parameters is that these parameters are thought to remain unchanged in such a short observation time and we are interested in the difference between the spectra of the two intervals. Other parameters settings are same as M13.

For our first step in spectral analysis, we checked the residuals by fixing the best-fit spectral parameters of interval~2 to the interval~1 spectrum. The residuals are shown in Figure~7. The $\bigtriangleup\chi$ value of the interval~1 spectrum is too high, indicating that the spectrum has changed markedly during the transition from interval~1 to interval~2. The main difference derives from the energy band below 10 keV (see Figure 7). The spectra of the two intervals are shown in Figure~8 and the best-fit parameters are shown in Table~2. We found that the power-law index increases from $1.66\pm0.01$ to $1.74\pm0.01$, suggesting that the spectra become soft during the transition. The emissivity break radius has a slight decrease from $6.31\pm0.12$ to $6.17\pm0.11$ $r_{\rm{g}}$.

\section{DISCUSSION}

In this paper, we have presented a timing analysis of \nustar\ observations of GRS 1915+105. We find a positive correlation between the QPO frequency and source flux (see Figure 5). The variable QPO frequency could be related to the inner disk edge\citep{Takizawa97}. If the count rate represents the mass accretion rate, this relationship implies that the movement of the inner disk edge may be driven by the change of mass accretion rate. During the evolution from low flux levels to high flux levels, the QPO frequency has increased from $\sim1.5$ Hz to $\sim1.85$ Hz, and the spectrum has become soft. The spectral parameters of the reflection component have no appreciable change during the transition (see Table 1), suggesting that the reflection component has little influence on the variation of QPO frequency.

GRS 1915+105 is the first galactic object to exhibit superluminal jets \citep{Mirabel94}, with the jets having primarily observed in the $\chi$ class \citep{Fender04}. Abundant QPOs with frequencies ranging from 0.5 to 10 Hz are observed in this system. It is suggested that these QPOs are associated with a corona \citep{Yan13}. The X-ray spectra of BHTs with a compact jet represent the sum of a thermal component emitted from the standard thin disk, a synchrotron emission component from the jet, Comptonization either through synchrotron self-Compton from the jet or inverse Compton from the corona, and a reflection component. \citet{Rodriguez04} found that the energy dependence of the QPO amplitude shows a turnover or flattening above 10 keV, which is similar to our results. Such relationship could be understood if the X-ray emission originates from two different radiation mechanisms: the X-ray emission above the break energy would be dominated by the synchrotron radiation from the jet, while the X-ray emission below the break energy would originate though Comptonization. Assuming that the QPO is contained in the Comptonized flux but not in the synchrotron flux, then the QPO amplitude would decrease in the high energy band (above the break energy) \citep{Rodriguez04}. As shown in Figure 3 and 7, we clearly find that the sudden jump in light curve and the spectral differences between the two intervals are all distinct in the energy bands below 10 keV. Given that the QPO frequency of the two intervals has changed, we can also consider that the 0.5-2 Hz QPO may be contained in the Comptonized flux below 10 keV which dominates the changes in light curve and energy spectrum. Based on the timing analysis of GRS 1915+105 in its $\rho$ class, \citet{Yan13} found that jet seems to have no influence on the LF-QPO frequency. Although we cannot eliminate the contribution of synchrotron self-Compton from the jet to the Comptonized flux, it is more likely that the 0.5-2 Hz QPO may originate from the corona. The centroid frequency of the 0.5-2 Hz QPO shows a negative correlation with photon energy \citep{Qu10}. In consideration of the relatively low frequency, these QPOs may relate to the outer part of the corona which region has a relatively lower density.

Among several models to explain the origin of LF-QPOs in black hole binaries, probably the most promising one relating QPOs to Comptonized flux from the corona is Lense-Thirring precession model proposed by \citet{Ingram09}. This model is based on the truncated disk assumption \citep{Done07}. The outer part is a cold, optical thick, geometrically thin disk truncated at some radius, and the inner part forms a hot, geometrically thick, optically thin accretion flow, which emits hard X-ray through inverse Compton scattering. QPOs arise from Lense-Thirring precession of the hot inner flow extending from an inner radius which larger than ISCO to the truncation radius. The variable QPO frequency could be associated with the movement of the truncation radius.

The results of spectral fits in this paper show that the inner emissivity index is extremely steep ($q=10$), while the outer emissivity profile becomes flat in both interval 1 and interval 2. Such steep inner emissivity index was also observed in some other black hole systems \citep{Fabian12}. This provides evidence that there are two different forms of accretion flows in the disk. The inner part is a hot, compact accretion flow emitting hard X-ray in the vicinity of black hole \citep{Wilkins11} and the outer part forms a cold flow with a flat emissivity index beyond the emissivity break radius. It is possible that the observed 0.5-2 Hz QPOs result from Lense-Thirring precession of the inner hot flow with a steep emissivity index. During the transition from interval 1 to interval 2, the emissivity break radius has a slight decrease accompanied with an increase in mass accretion rate if we can regard count rate as a proxy of mass accretion rate (see Table 2). Then, the increase in QPO frequency can be interpreted as the movement of the break radius: the Lense-Thirring precession frequency is $\upsilon_{\rm{LT}}=GMa/{\pi}c^{2}r^{3}$ \citep{Stella98}. When mass accretion rate has a rapid increase, the outer disk gradually moving inward leads to an increase in Lense-Thirring precession frequency. This scenario can also explain spectral changes during the transition. As mass accretion rate increases in the LHS, the source evolves along the right vertical branch from bottom to top in the HID and the energy spectrum is dominated by hard component. After reaching the peak, it experiences a transition to the HIMS corresponding to the horizontal branch, and the hardness gradually decreases with the emergence of disk component \citep{Belloni11}. The inward movement of the outer disk with the increase in mass accretion rate results in a stronger disk component and greater cooling of the hot flow by the cool disk photons and hence a softer spectrum.

A similar conclusion was reported by \citet{Sriram13} with the \rxte\ observations. They used the normalization of the diskbb model for a measurement of the inner disk radius, and found that the movement of the coupled inner disk-corona region leads to the change in QPO frequency. However, this approach to measure the inner disk radius is not accurate and the measured inner disk radius is always underestimated \citep{Merloni00}. Though the emissivity break radius changes marginally in our spectral fits, it still gives a more direct evident for the movement of the break radius.

The global disk oscillations (GDM) model could also explain observed $\sim1$ Hz QPOs in some black hole transients \citep{Titarchuk00}. In this model, the QPOs are caused by vertical oscillations of the disk as a whole body and the gravitational interaction between the central black hole and the disk plays an important role on the formation of the oscillations. However, the GDM frequency tends to a persistent frequency because the oscillations are characteristic of the local properties of the disk and magnetic fields, and hence can not vary in a short time. Moreover, the model expects that the QPO frequency is not correlated with the source flux, which is in contrast to our results.

In summary, we observe a type-C QPO below 2 Hz in the \nustar\ data of GRS 1915+105. The QPO shows a sudden increase in frequency along with a source flux increase. After further study, we find that the QPO frequency is correlated with the 3-25 keV count rate. During the transition from low-flux level to high-flux level, spectral properties have obvious variations. It is shown a spectral softening and a decreased trend of truncation radius. Our results indicate that reflection component has little effect on the QPO frequency, and the change in QPO frequency may be associated with the movement of the outer part of the disk.

\acknowledgements
We thank H. Q. Gao, Z. Zhang for theory suggestions. This research has made use of data from the {\it NuSTAR} mission, a project led by the California Institute of Technology, managed by the Jet Propulsion Laboratory, and funded by the National Aeronautics and Space Administration. We thank the {\it NuSTAR} Operations, Software and Calibration teams for support with the execution and analysis of these observations.  This research has made use of the {\it NuSTAR} Data Analysis Software (NuSTARDAS) jointly developed by the ASI Science Data Center (ASDC, Italy) and the California Institute of Technology (USA). This work is supported by the National Basic Research program of China 973 Program 2009CB824800, the National Natural Science Foundation of China (11173024) and the Fundamental Research Funds for the Central Universities, the Strategic Priority Research Program on Space Science, the Chinese Academy of Sciences, Grant No. XDA04010300.

\clearpage


\begin{figure}
  \centering
  \includegraphics[width=90mm]{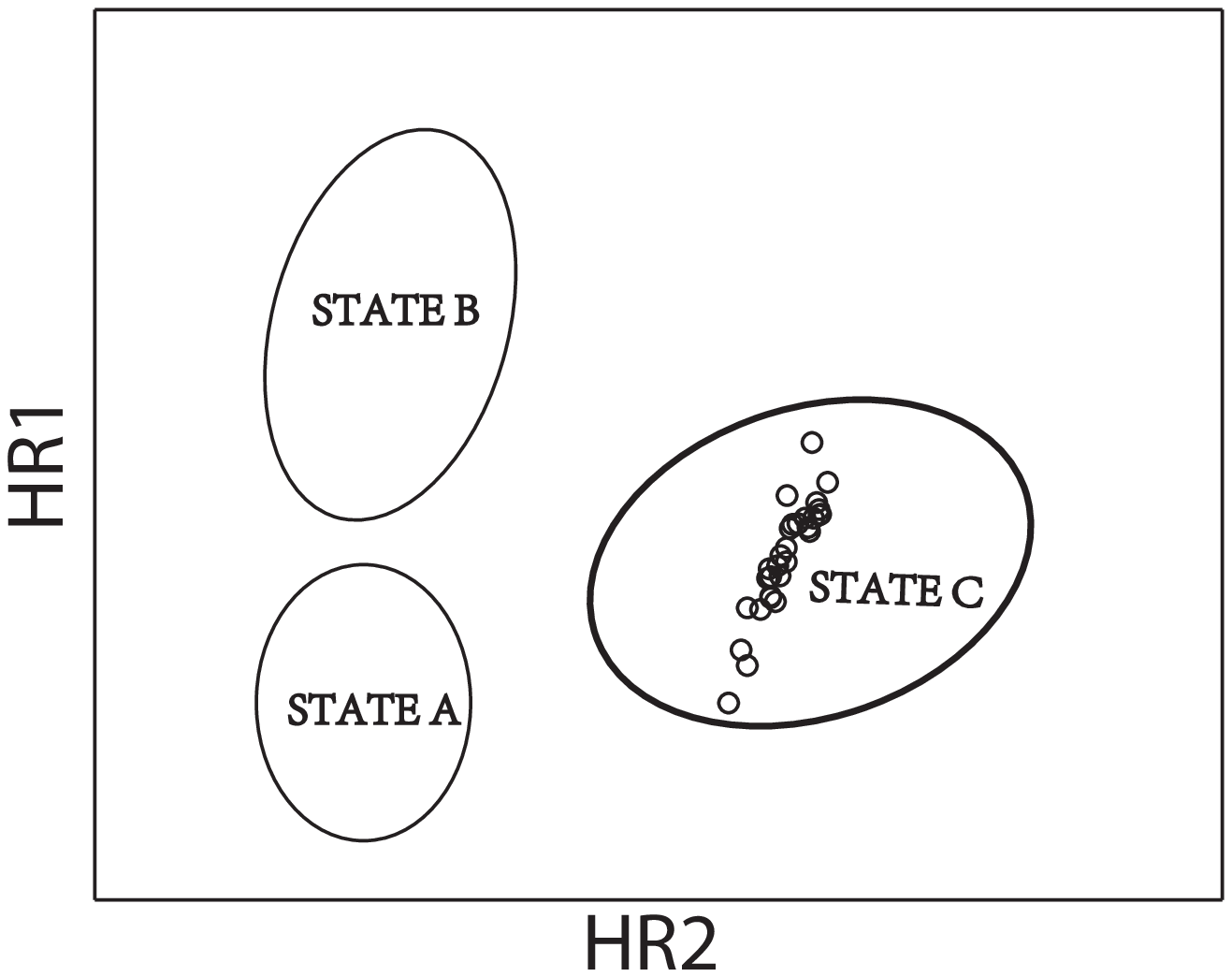}
  \caption{A schematic color-color diagram shows the basic A/B/C states of GRS 1915+105. Figure adapted from \citet{Belloni00}. The black data points in state C are the \nustar\ data. HR1=(7-15 keV)/(3-7 keV) and HR2=(15-79 keV)/(3-7 keV).}
  \label{fig1}
\end{figure}

\begin{figure}
 \centering
 \epsscale{.8}
 \plotone{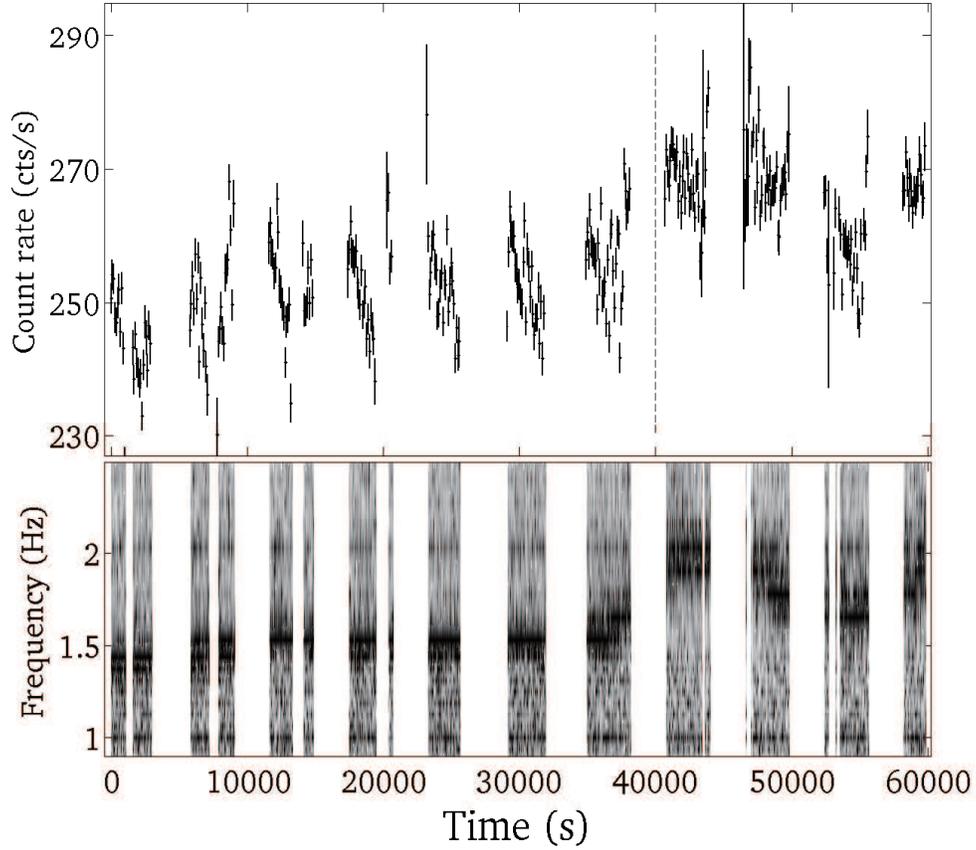}
 \caption{Top: The 3-25 keV \nustar\ FPMA light curve of GRS 1915+105. The dash line separates the two intervals used for obtaining the PDS and spectra (see the text). Bottom: Dynamical power density spectrum in the same band, taken at 16~s intervals. Note that the signals at multiples of 1 Hz are due to housekeeping operations. In PDS fits, we removed these data points.}
 \label{fig2}
\end{figure}

\begin{figure}
  \centering
  \includegraphics[width=90mm]{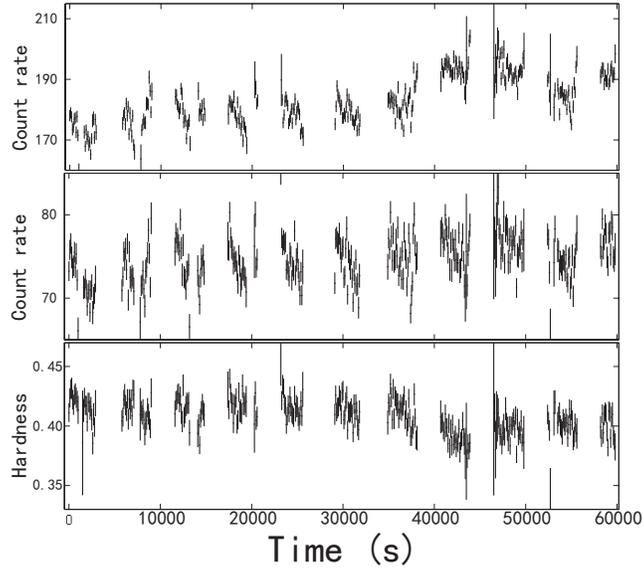}
  \caption{The 3-9 keV light curve (top), 9-25 keV light curve (middle) and hardness (9-25 keV/3-9 keV) (bottom) of GRS 1915+105 from the \nustar\ FPMA.}
  \label{fig3}
\end{figure}

\begin{figure}
  \centering
  \includegraphics[width=90mm]{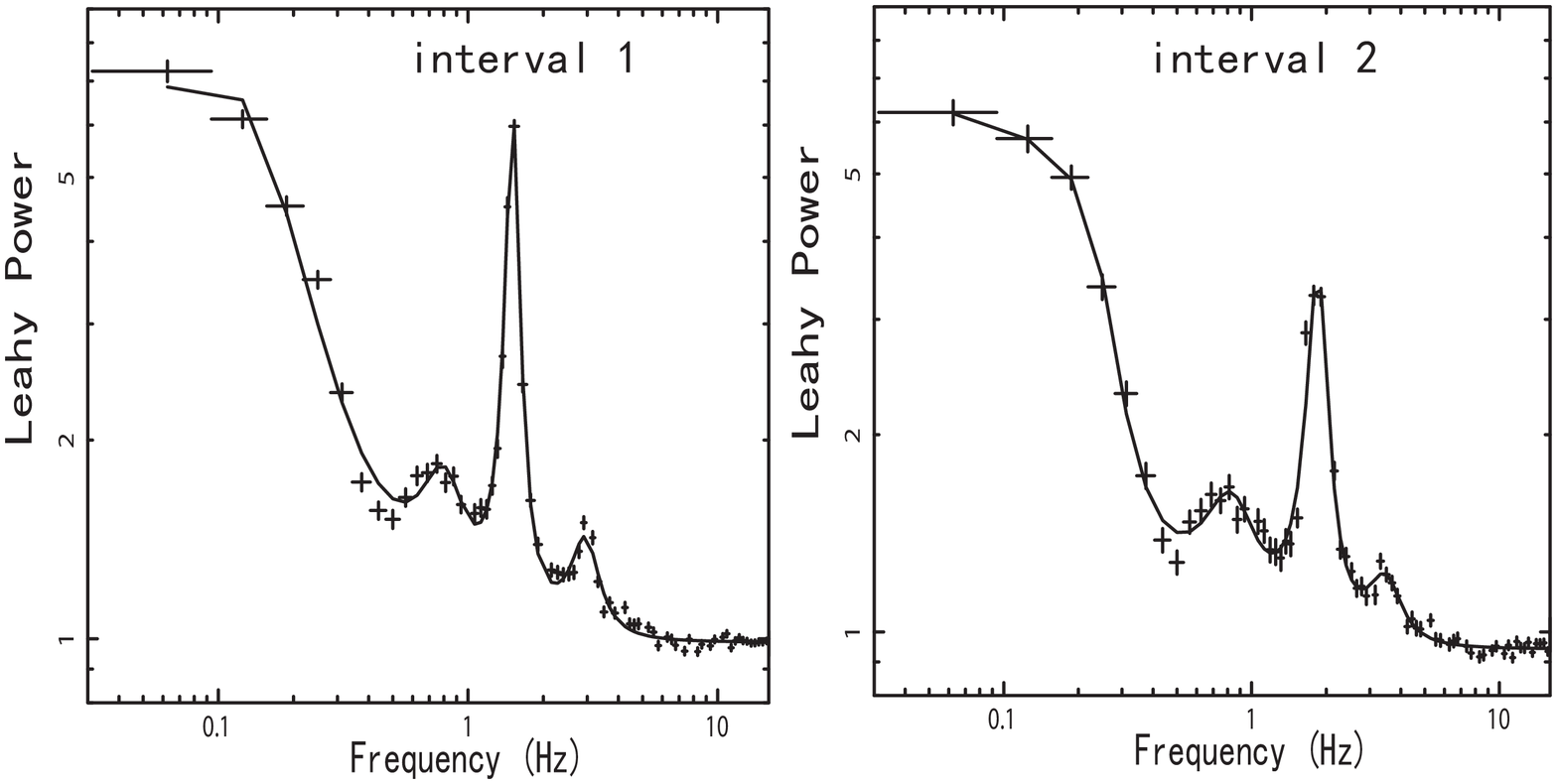}
  \caption{Power Density Spectra of interval 1 (left) and interval 2 (right) in the 3-25 keV band.}
  \label{fig4}
\end{figure}

\begin{figure}
  \centering
  \includegraphics[width=90mm]{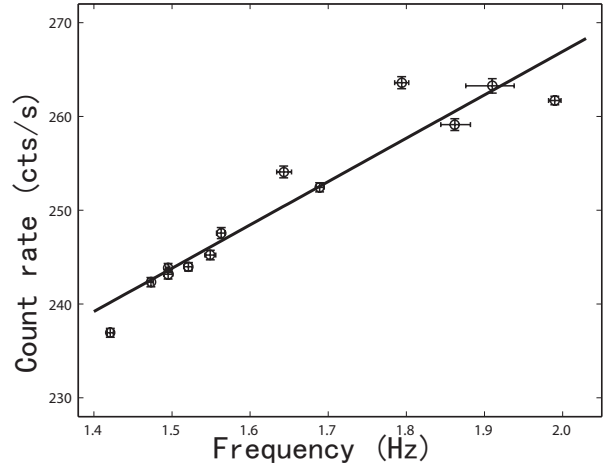}
  \caption{QPO centroid frequency vs. 3-25 keV count rate.}
  \label{fig5}
\end{figure}

\begin{figure}
  \centering
  \includegraphics[width=120mm]{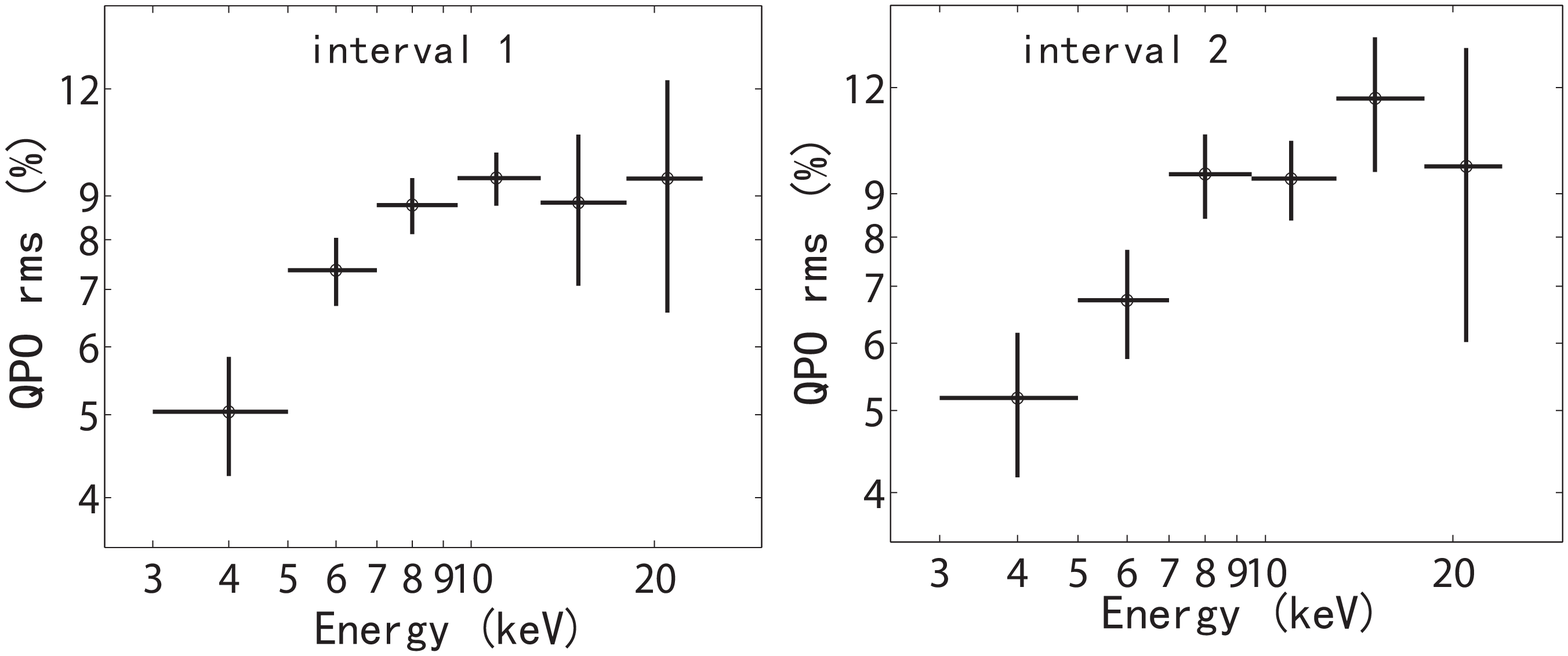}
  \caption{The energy dependence of QPO rms amplitude for interval 1 (left) and interval 2 (right).}
  \label{fig6}
\end{figure}

\begin{figure}
  \centering
  \includegraphics[width=90mm]{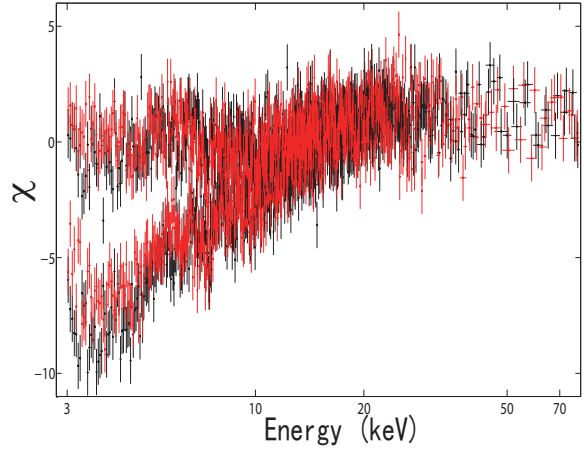}
  \caption{Residuals of interval 1 and interval 2 by fixing the best-fit spectral parameters of interval 2 to the interval 1 spectrum. Black and red represent FPMA and FPMB, respectively.}
  \label{fig7}
\end{figure}

\begin{figure}
  \centering
  \includegraphics[width=90mm]{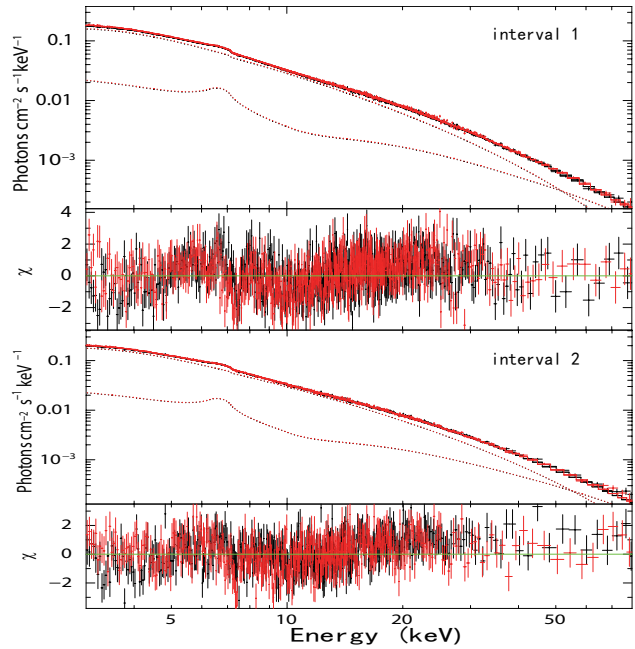}
  \caption{FPMA (black) and FPMB (red) spectra of interval 1 (top) and interval 2 (bottom), using a relativistically blurred disk reflection model.}
  \label{fig8}
\end{figure}






\clearpage

\begin{table}
\begin{center}
\caption{QPOs properties of interval 1 and interval 2.}
\begin{tabular}{@{}ccccccc}
\tableline\tableline
QPO components & \multicolumn{3}{c}{interval 1} & \multicolumn{3}{c}{interval 2}\\
 \tableline
 & $\upsilon_{0}$ & rms & $Q$ & $\upsilon_{0}$ & rms & $Q$\\
 & (Hz) & (\%) & & (Hz) & (\%) &\\
\tableline
fundamental & $1.51\pm0.01$ & $7.46\pm0.21$ & 8.42 & $1.84\pm0.01$ & $8.04\pm0.59$ & 4.55\\
second harmonic & $2.94\pm0.02$ & $4.81\pm0.40$ & 3.23 & $3.45\pm0.05$ & $4.30\pm0.49$ & 2.74\\
subharmonic & $0.79\pm0.01$ & $3.83\pm0.30$ & 2.03 & $0.81\pm0.02$ & $3.97\pm0.37$ & 1.18\\
\tableline
\end{tabular}
\end{center}
\end{table}

\begin{table}
\begin{center}
\caption{Best-fit parameters for the interval 1 and interval 2 spectra.}
\begin{tabular}{@{}cccccccccc}
\tableline\tableline
Model & $q_{\rm{in}}$ & $q_{\rm{out}}$ & $r_{\rm{break}}$ & $\Gamma$ & $E_{\rm{cut}}$ & $K_{\rm{pow}}$ & $\xi$ & $K_{\rm{refl}}$ & $\chi^{2}/\upsilon$\\
 &  & & $(r_{\rm{g}})$ & & (keV) & & (erg cm $\rm{s}^{-1}$) & ($10^{-5}$) & \\
\tableline
interval 1 & 10 & 5.50E-06 & 6.31 & 1.66 & 22.72 & 2.12 & 999.8 & 1.43 & 4428.2/3798\\
interval 2 & 10 & 9.56E-07 & 6.17 & 1.74 & 23.54 & 2.56 & 991.4 & 1.40 & 3876.0/3798\\
\tableline
\end{tabular}
\tablecomments{$q_{\rm{in}}$ and $q_{\rm{out}}$ are the inner and outer disk emissivity indices, respectively. $r_{\rm{break}}$ is the emissivity break radius, $\Gamma$ is the photon index of the cut-off power law, and $\xi$ is the ionization parameter. The cut-off power-law normalization, $K_{\rm{pow}}$, has units of photons $\rm{cm}^{-2}$ $\rm{s}^{-1}$ $\rm{keV}^{-1}$ at 1 keV. $E_{\rm{cut}}$ is the cut-off energy, and $K_{\rm{refl}}$ is the reflection normalization. }
\end{center}
\end{table}

\end{document}